\documentclass[a4paper]{article}
\usepackage{multirow}

\usepackage{INTERSPEECH2021}

\title{The DKU-DukeECE Diarization System for \\ the VoxCeleb Speaker Recognition Challenge 2022}
\name{Weiqing Wang$^{2}$, Xiaoyi Qin$^{1}$, Ming Cheng$^1$, Yucong Zhang$^1$, Kangyue Wang$^1$, Ming Li$^{1, 2}$}

\address{
  $^1$ Data Science Research Center, Duke Kunshan University, Kunshan, China \\
  $^2$ Department of Electrical and Computer Engineering, Duke University, Durham, USA
  }
\email{ming.li369@duke.edu}

\begin{document}

\maketitle

\begin{abstract}
This paper discribes the DKU-DukeECE submission to the 4th track of the VoxCeleb Speaker Recognition Challenge 2022 (VoxSRC-22). Our system contains a fused voice activity detection model, a clustering-based diarization model, and a target-speaker voice activity detection-based overlap detection model. Overall, the submitted system is similar to our previous year's system in VoxSRC-21. The difference is that we use a much better speaker embedding and a fused voice activity detection, which significantly improves the performance. Finally, we fuse 4 different systems using DOVER-lap and achieve 4.75\% of the diarization error rate, which ranks the 1st place in track 4. 
\end{abstract}

\section{Introduction}

Speaker diarization is the task that breaks up multi-speaker audio into homogeneous single speaker segments, effectively solving ``who spoke when'', which is also the task in track 4 of the VoxCeleb Speaker Recognition Challenge 2022 (VoxSRC-22). In this paper, we focus on the speaker diarization tasks and present the details of our submitted system.

Compare with our system in last year's VoxSRC (VoxSRC-21)~\cite{myVoxSRC21, VoxSRC21}, we focus more on the voice activity detection (VAD) model this year. We tried several different VAD models. Although each of the single VAD model only shows similar performance compared with our previous year's submission, but the fusion brings a satisfying reduction on diarization error rate (DER). In addtion, for overlap detection, the target-speaker voice activity detection (TS-VAD) shows better performance than just only predicting the overlap regions. Last but not least, we employ a more powerful speaker embedding model~\cite{qinxy_simam} in the agglomerative hierarchical clustering-based (AHC) diarization model. This AHC model with TS-VAD can achieve 4.85\% of the DER, which still ranks 1st on the leaderboard. Finally, fusing of 4 different systems with DOVER-lap further reduces the DER to 4.74\%.

\section{Dataset Description}

As this task can use any dataset for training, the detailed dataset used in this challenge for each model include:

\begin{itemize}

\item Voice activity detection (VAD): %DIHARD II~\cite{DIHARDII}, DIHARD III~\cite{DIHARDIII} and 
Voxconverse dev set for training and Voxconverse test set for validation. 
\item Speaker embedding: Voxceleb 1 \& 2~\cite{Voxceleb} for training.
\item AHC-based diarization system: Only Voxconverse test set for hyper-parameter tuning.
\item Spectral clustering-based diarization system: Voxconverse dev set for training and Voxconverse test set for validation. 
\item Overlap detection: Voxconverse dev set for training and Voxconverse test set for validation. 
\item TS-VAD: This model is first trained on the data simulated by Librispeech~\cite{Librispeech}. Then it is finetuned on Voxconverse dev set and validated on Voxconverse test set.
\item Data augmentation: MUSAN~\cite{MUSAN} and RIRs~\cite{RIRs} corpus.
\end{itemize}

\section{Detailed Model Configuration}
This section describes the model of our submission. If not specified, the input acoustic features of all model are 80-dim log Mel-filterbank energies with a frame length of 25ms and a frameshift of 10ms. 

\subsection{VAD}
We focus more on the VAD in this year's challenge, and we employ 4 different models for VAD.
\subsubsection{ResNet34-based VAD (model \#1)}
We use a ResNet34 as the front-end model to extract the frame-level feature map. Next, a statistical pooling layer is employed on the feature map every one frames. Finally, a 4-head 2-layer transformer encoder and a fully-connected layers with a sigmoid function produce the posterior probability of speech. 

\subsubsection{Conformer-based VAD (model \#2)}
We adopt a ResNet50 as the front-end model to extract multi-scale feature maps as same as the configuration in \cite{zhang2022dino}. Then, the convolution subsampling layers transfer each feature map to a sequence of embeddings. Next, a Conformer encoder \cite{gulati20_interspeech} aggregates the flattened feature sequence and the vanilla Transformer decoder \cite{vaswani2017attention} can predict the final VAD results. The encoder and decoder are set to have 6 layers with 256-dim 4-head attention modules and sinusoidal positional encodings.

\subsubsection{VAD from \textit{pyannote.audio 2.0} (model \#3)}
We use \textit{pyannote.audio 2.0}\footnote{https://github.com/pyannote/pyannote-audio/tree/develop} for computing the VAD results. We use the same model and hyper-parameters hosted in \textit{huggingface}\footnote{https://huggingface.co/pyannote/segmentation}.

\subsubsection{ASR-based VAD (model \#4)}
We make use of the word-level timestamps derived from the Kaldi ASR system. Intuitively, all the time segments labeled with \textit{eps} symbols are regarded as non-speech segments. In this challenge, we follow the same recipe provided in Kaldi Librispeech recipe\footnote{https://kaldi-asr.org/models/m13}~\cite{povey2011kaldi}, except that we change the window shift to 12.5 milliseconds when computing the MFCCs.

\begin{table}[htb]
  \caption{False alarm (FA), miss detection (MISS) and accuracy of the VAD model on Voxconverse test set}
  \label{tab:vad}
  \centering
  \begin{tabular}[c]{cccc}
    \toprule
    \textbf{\#Model}  & \textbf{FA [\%]} & \textbf{MISS [\%]} & \textbf{Error} [\%] \\
    \midrule
    1      & 2.94 & 1.33 & 4.27 \\
    2      & 2.70 & 1.77 & 4.47 \\
    3      & 2.25 & 2.10 & 4.35 \\
    4      & 0.81 & 11.87 & 12.68 \\
    Fusion & 2.60 & 1.37 & 3.97 \\
    \bottomrule
  \end{tabular}
\end{table}

As shown in Table \ref{tab:vad}, model 1, 2 and 3 have similar performance. The performance of model 4 is worse than others as it is a pretrained ASR system and we cannot tune it on the voxconverse dataset without text labels. After we perform fusion with majority vote, both false alarm and miss are reduced, which brings about 0.3\% reduction on DER.

\subsection{Speaker Embedding}
The SimAM-ResNet34 structure is employed as the front-end pattern extractor, which learns a frame-level representation from the input acoustic feature. An attentive statistic pooling (ASP) \cite{asp_pooling} layer projects the variable length input to the fixed-length vector. Next, a 256-dim fully connected layer is adopted as the speaker embedding layer. The ArcFace (s=32,m=0.2) \cite{arcface} is used as a classifier. The detailed configuration of the neural network is the same as \cite{qinxy_simam}. The acoustic feature is 80-dim log Mel-filterbank energies with a frame length of 25ms and a frameshift of 10ms. 

We adopt the two-stage training method. The speaker embedding model in the pre-trained stage is trained by VoxCeleb2 dev set. To verify the performance of the speaker verification system in the VoxSRC22 task4, we create the trial file using relative speaker labels in segments (since there are no ground truth speaker labels in the VoxConverse set). Table. \ref{tab:asv_result} reports the system performance. The results indicate a large domain gap between the VoxCeleb and VoxConverse sets. Therefore, to reduce the domain mismatch, the embedding of VoxConverse dev set is extracted and clustered with a fixed threshold to determine the speaker class. After pseudo-labeling for the VoxConverse dev set, the Vox2dev set together with the VoxConverse dev set are fed into the pre-trained speaker model to fine-tune. 

However, although finetuned embedding shows better performance on voxconverse test set, but it is worse on the challenge test set. Therefore, we only present the results that use the embedding trained by VoxCeleb2.

\begin{table}[tp]
  \caption{ The performance of speaker embedding system.}
  \label{tab:asv_result}
  \centering
  \begin{tabular}[c]{lcccccccc}
    \toprule
    \multirow{2}*{ Model } & \multicolumn{2}{c}{Vox-O} & \multicolumn{2}{c}{VoxSRC22 task4val }   \\ 
    \cmidrule(lr){2-3} \cmidrule(lr){4-5} & EER[\%] & mDCF & EER[\%] & mDCF \\
   \midrule
   SimAM-ResNet  & 0.726 & 0.036 & 5.84 & 0.220\\
   \quad + fine-tune  & - & - & 5.08 & 0.335 \\
     \bottomrule
     \end{tabular}
\end{table}

\subsection{Clustering-based Diarization}
For clustering-based diarization, we use two different model. One is based on AHC and another is based on spectral clustering. In our experiments, AHC can achieve a lower DER and spectral clustering can estimated the number of speaker more accurately.
\subsubsection{AHC}
The AHC-based diarization model is exactly the same as we used in previous years, which is also similar to the Microsoft system in VoxSRC-20 without speech separation~\cite{microsoft}. First, speaker embeddings are extracted from the uniformly segmented speech with a length of 1.28s and shift of 0.32s, and two consecutive segments are merged into a longer segment if the distance is greater than a segment threshold, which is the AHC-based segmentation. The pairwise similarity is measured by cosine distance. Next, we perform a plain AHC on the similarity matrix with a relatively high stop threshold to obtain the clusters with high confidence. These clusters are split into ``long clusters'' and ``short clusters'' by the total duration in each cluster, and the central embedding of each cluster is the mean of all speaker embeddings in the cluster. Later, each short cluster is assigned to the closest long cluster by the cosine distance of central embedding. Finally, if a short cluster is too different from all long clusters, which means that the distance between them is lower than a speaker threshold, we treat it as a new speaker. 

All of these parameters are directly tuned on the voxconverse test set by grid search. In our experiments, the segment threshold is 0.54, the stop threshold is 0.6, the duration for classifying long and short clusters is 6s, and the speaker threshold is 0.2. 

\subsubsection{Spectral Clustering}
We use an LSTM model to predict the affinity matrix~\cite{LinLSTM}. The model consists of two BiLSTM and two fully connected layers with a sigmoid function. Speaker embeddings are also extracted from the uniformly segmented speech with a length of 1.28s and shift of 0.64s. Next, the speaker embedding sequence $[\mathbf{x}_1, \mathbf{x}_2, ..., \mathbf{x}_n]$ is concatenated with repeated $\mathbf{x}_i$ as the input and produce the i-th row of the affinity matrix $\mathbf{S}$:

\begin{equation}
    \label{eq:BiLSTM}
    \mathbf{S}_i = [\mathbf{S}_{i,1}, \mathbf{S}_{i,2}, ... , \mathbf{S}_{i,n}] = f(
    \left[\begin{matrix}\mathbf{x}_i\\\mathbf{x}_1\end{matrix}\right], 
    \left[\begin{matrix}\mathbf{x}_i\\\mathbf{x}_2\end{matrix}\right], ...,
    \left[\begin{matrix}\mathbf{x}_i\\\mathbf{x}_n\end{matrix}\right]),
\end{equation}
where $f$ is the LSTM-based neural network, n is set to 64 in our experiments. More details can be found in~\cite{DIHARDII-LSTM} and~\cite{wang2022similarity}. 

The model is trained on the voxconverse dev set for 200 epochs. The model is optimized with BCE loss and Adam optimizer with a learning rate of 0.001. After obtaining the affinity matrix $\mathbf{S}$, we employ spectral clustering (SC) to get the final diarization results. 

\subsection{Overlap Detection}
The overlap detection model is the same as that of the VAD model \#1. The label is 1 for overlapped speech and 0 otherwise. After an overlapped region is detected, we replace the label with two closest speakers near this region. The threshold for overlap decision is set to 0.85. The input is 16s chunked wav.

\begin{table*}[tp]
  \caption{ The performance of different speaker diarization systems in terms of DER (\%) and JER (\%).}

  \label{tab:result}
  \centering
  \begin{tabular}[c]{lcccccccc}
    \toprule
     \multirow{2}*{\textbf{Model}} & \multicolumn{2}{c}{\textbf{Test (Oracle VAD)}} &  \multicolumn{2}{c}{\textbf{Test (System VAD)}} & \multicolumn{2}{c}{\textbf{VoxSRC-22 Test}}\\
      \cmidrule(lr){2-3} \cmidrule(lr){4-5} \cmidrule(lr){6-7} 
     
      & \textbf{DER[\%]}  & \textbf{JER[\%]} & \textbf{DER[\%]}  & \textbf{JER[\%]} & \textbf{DER[\%]}  & \textbf{JER[\%]} &\\
 
   \midrule
   Baseline                                 & - &   -  &  -  & -  & 19.60 & 41.43 \\
   \midrule
	AHC                                     & 3.36 & 21.67 & 5.35 & 27.99 & - & -  \\
	\ \ + \textit{OD}                       & 3.03 & 21.43 & 5.02 & 27.72 & - & -  \\
	\ \ + \textit{TS-VAD (fully assigned)}    & 3.60 & 22.21 & 5.61 & 28.08 & - & -  \\
	\ \ + \textit{TS-VAD (partially assigned)}& 2.96 & 21.77 & 4.86 & 27.69 & 4.85 & 28.05  \\

	LSTM-SC                                 & 4.91 & 32.74 & 6.36 & 34.82 & - & -  \\
    \ \ + \textit{OD}                       & 4.39 & 32.02 & 6.04 & 34.53 & - & -  \\
	\ \ + \textit{TS-VAD (fully assigned)}    & 4.12 & 31.70 & 5.68 & 33.92 & - & -  \\
	\ \ + \textit{TS-VAD (partially assigned)}& 4.31 & 32.14 & 5.85 & 34.30 & - & -  \\
    
	 \midrule
	 Fusion & 3.09 & 23.14 & 4.94 & 28.79 & 4.74 & 27.84  \\
	  
     \bottomrule
     \end{tabular}
\end{table*}

\subsection{TS-VAD}
\label{sec:TSVAD}

\subsubsection{Data Simulation}
we create a simulated dataset from the Librispeech dataset, and the simulation process is as follows: 
\begin{enumerate}
    \item We select all non-overlapped speech for each speaker from the Librispeech dataset for simulation. 
    \item Extract the labels from the transcript of the voxconverse dev set and remove all silence regions. 
    \item During the training stage, the simulated data is generated in an online manner, where we randomly choose a segment of the label and fill the active region with the continuous non-overlapped speech segments.
\end{enumerate}
The more detailed simulation process can be found in~\cite{myVoxSRC21}. Finally, the voxconverse test set is adopted as the validation and evaluation set. 

\subsubsection{Training details}
TS-VAD has achieved an excellent performance on CHIME6~\cite{tsvad} and DIHARD III~\cite{dihard3tsvad} challenge. Unlike the previous method using i-vector, we employ ResNet-based x-vector as the target-speaker embedding. 

The TS-VAD model is also similar to the VAD model except that the feature maps produced by ResNet need to be concatenated with a target speaker embedding. The concatenated features are then fed to the BiLSTM layers and fully connected layers. The number of target speakers embedding $N$ is set to 8. The parameters of front ResNet34 are initialized from another ResNet34-based speaker embedding model. 

The model is first trained on the simulated LibriSpeech for 10 epochs with front ResNet34 frozen, and then it is trained for another 10 epochs with all parameters. Finally, we fine-tune the model on voxconverse dev set for 50 epochs and validate on voxconverse test set. The learning rate is 0.0001 when training on simulated data and 0.00001 during the fine-tuning stage. The model is optimized by BCE loss and Adam optimizer. The input is 16s chunked wav. 

\subsubsection{Inference}
For inference, the non-speech regions are first removed by VAD, and the wavs are split into 16s chunks. Next, speaker embeddings are extracted given the results from a clustering-based method. We only consider those speaker embeddings with 16s or longer speech. For those speakers whose speech is shorter than 16s, we directly keep their clustering-based results. If the number of speakers is less than 8, we use zero-vectors as the fake embeddings. If it is greater than 8, we discard the speaker embeddings with shorter speech, but their labels are kept in the final results. The threshold for speaker decision is set to 0.5. 

However, if we fully assign the TS-VAD results for each speaker, the performance is worse compared with the AHC-based results as the retrieved overlap regions cannot compensate the error on speaker confusion. To solve this, we only partially assign the overlap region to the AHC-based results, which significantly reduces the DER.

\subsection{Data Augmentation}
We perform online data augmentation~\cite{cai2020fly} with MUSAN and RIRs corpus. For background additive noise, we use ambient noise, music, television, and babble noise. For reverberation, we perform convolution with 40,000 simulated room impulse responses from small and medium rooms. The data augmentation is employed for all models which take acoustic features as input. 

\subsection{System Fusion}
To further improve the performance and robustness, we fuse our systems by DOVER-Lap~\cite{doverlap}. 

\section{Experimental Results}

Table \ref{tab:result} shows the results on voxconverse test set and the challenge test set. For the clustering-based system, the AHC method achieves $5.35\%$ of DER on Voxconverse test set with system VAD. Next, we use TS-VAD to detection the overlap regions and achieves $4.85\%$ of DER on both Voxconverse test set and challenge test set. 

Although AHC-based diarization system can achieve the lowest DER, but it always overestimates the number of speaker, e.g., predicting 30+ speakers even it only contains 10 speakers. To solve this, we employ SC-based diarization system. Although it shows worse performance compared with the AHC-based system, it can always estimates the number of speaker more accurately than AHC-based system. We didn't submit this SC-based system independently, but we fuse it with other systems to improve the system diversity and achieve a lower DER. 

Finally, our best system contains 4 systems fused by DOVER-Lap: AHC+OD, AHC+TS-VAD (partially assigned), AHC+TS-VAD (fully assigned) and LSTM-SC+TS-VAD (partially assigned), which achieves $4.75\%$ of the DER, which ranks 1st place in this challenge. 

\section{Conclusions}

In this paper, we describe our system for the VoxSRC 2022. To achieve better performance, we mainly focus on the voice activity detection and overlapped speech detection. We employ overlap detection and TS-VAD to reduce the missed speaker error. Our fused VAD reduces the DER by $0.3\%$ compared with the VAD from single model, and the TS-VAD-based overlap detection further reduces DER by about $0.6\%$, which significantly improves the performance. 

\bibliographystyle{IEEEtran}

\bibliography{mybib}

% \begin{thebibliography}{9}
% \bibitem[1]{Davis80-COP}
%   S.\ B.\ Davis and P.\ Mermelstein,
%   ``Comparison of parametric representation for monosyllabic word recognition in continuously spoken sentences,''
%   \textit{IEEE Transactions on Acoustics, Speech and Signal Processing}, vol.~28, no.~4, pp.~357--366, 1980.
% \bibitem[2]{Rabiner89-ATO}
%   L.\ R.\ Rabiner,
%   ``A tutorial on hidden Markov models and selected applications in speech recognition,''
%   \textit{Proceedings of the IEEE}, vol.~77, no.~2, pp.~257-286, 1989.
% \bibitem[3]{Hastie09-TEO}
%   T.\ Hastie, R.\ Tibshirani, and J.\ Friedman,
%   \textit{The Elements of Statistical Learning -- Data Mining, Inference, and Prediction}.
%   New York: Springer, 2009.
% \bibitem[4]{YourName17-XXX}
%   F.\ Lastname1, F.\ Lastname2, and F.\ Lastname3,
%   ``Title of your INTERSPEECH 2021 publication,''
%   in \textit{Interspeech 2021 -- 20\textsuperscript{th} Annual Conference of the International Speech Communication Association, September 15-19, Graz, Austria, Proceedings, Proceedings}, 2020, pp.~100--104.
% \end{thebibliography}

\end{document}